\documentclass[aps,twocolumn,showpacs,floatfix,superscriptaddress,showkeys]{revtex4}
\usepackage{graphicx}
\usepackage{psfrag}
\begin{document}
\newcommand{\beq}{\begin{equation}}
\newcommand{\eeq}{\end{equation}}
\newcommand{\ben}{\begin{eqnarray}}
\newcommand{\een}{\end{eqnarray}}
\newcommand{\bea}{\begin{array}}
\newcommand{\eea}{\end{array}}
\newcommand{\om}{(\omega )}
\newcommand{\bef}{\begin{figure}}
\newcommand{\eef}{\end{figure}}
\newcommand{\leg}[1]{\caption{\protect\rm{\protect\footnotesize{#1}}}}
\newcommand{\ew}[1]{\langle{#1}\rangle}
\newcommand{\be}[1]{\mid\!{#1}\!\mid}
\newcommand{\no}{\nonumber}
\newcommand{\etal}{{\em et~al }}
\newcommand{\geff}{g_{\mbox{\it{\scriptsize{eff}}}}}
\newcommand{\da}[1]{{#1}^\dagger}
\newcommand{\cf}{{\it cf.\/}\ }
\newcommand{\ie}{{\it i.e.\/}\ }
\title{Adiabaticity Conditions for Volatility Smile in Black-Scholes Pricing Model}
\author{L.~Spadafora}  
\affiliation {Dipartimento di Matematica e Fisica,
Universit\`a Cattolica, via Musei 41, 25121 Brescia, Italy}
\author{G.~P.~Berman}
\affiliation {Theoretical Division, MS-B213, Los Alamos National Laboratory, Los Alamos, NM, 87545}
\author{F.~Borgonovi}
\affiliation {Dipartimento di Matematica e Fisica,
Universit\`a Cattolica, via Musei 41, 25121 Brescia, Italy}
\affiliation{I.N.F.N. Sezione di Pavia, Pavia, Italy}
\begin{abstract}

Our derivation of the distribution function for future returns is based on the risk neutral approach which gives a functional dependence for the European call (put)
option price, $C(K)$, given the strike price, $K$, and the distribution function of the returns.
We derive this distribution function  using for $C(K)$ a Black-Scholes (BS) expression with volatility, $\sigma$, in the form of a
volatility smile. We show that this approach based on a volatility smile leads to relative minima for the distribution function (``bad" probabilities) never observed in real data and, in the worst cases, negative probabilities.
We show that these undesirable effects can be eliminated by requiring   ``adiabatic" conditions on the volatility smile. 
\end{abstract}

\date{\today}
\pacs{05.10.Gg, 05.40.Jc, 02.50.Le, 89.65.Gh }
\keywords{Volatility smile, Black-Sholes model, no-arbitrage conditions}

\maketitle


\section{Introduction}


One of the simplest  ``products" on the derivative financial market is the European
call (put) option~\cite{Hull, Wilmott}.
Considering the risk neutral approach, the price of the European call option, $C \equiv  C(S_T, K, T, r)$,
is defined by
\begin{equation}\label{pricing_risk_neutral}
C = e^{-rT} \int_{K}^{\infty} (S_T - K) P(S_T) d S_T,
\end{equation}
where $S_T$ is the stock price at time $t = T$, $K$ is the strike price of the option, $T$ is the 
expiration time (time to maturity) of the option, $r$ 
is the interest rate and $P(S_T) \ge 0$ is the distribution function of the stock prices in a ``risk-neutral world" ($\int_{0}^{\infty} P(S_T) d S_T =1$).

Eq.~(\ref{pricing_risk_neutral}) is too
general since it does not place any restrictions on the underlying stock price distribution
function, $P(S_T)$. To calculate explicitly the option price, $C$, using Eq.~(\ref{pricing_risk_neutral}), one must know
the distribution function, $P(S_T)$. Consequently, one must
 make some assumptions about
the stock prices.
An important achievement in the theory of option pricing is the Black-Scholes (BS)
theory which gives analytic solutions for the European call and put options~\cite{Black}.

In particular, for the European call option, a solution of the BS equation is given by Eq.~(\ref{pricing_risk_neutral}), if
one assumes for the distribution function, $P(S_T)$, a log-normal distribution,

\begin{equation}\label{normal_dist}
P(x) = \frac{1}{\sqrt{2\pi \sigma^2 (T - t)}} 
\exp\left[-\frac{(x + \sigma^2(T-t)/2)^2}{2 \sigma^2 (T-t)} \right],
\end{equation}
where $x = \ln(S_T/S(t)) - r(T-t)$ is the logarithmic return deprived of the risk-free 
component, $S(t)$  is the stock price at time $t$ and 
$\sigma$ is the stock price volatility. For seek of simplicity, in the following we consider $t = 0$ and we define $S_0 \equiv S(t=0)$.
Substituting Eq.~(\ref{normal_dist}) in (\ref{pricing_risk_neutral})  an explicit expression 
for the price of the European call option which satisfies the BS equation~\cite{Black} is obtained,
\begin{equation}\label{BS_solution}
C^{BS} = S_0 N(d_1) - K e^{-rT}N(d_2),
\end{equation}
where
\begin{equation}
\begin{array}{ccc}
d_1 &=& \displaystyle\frac{\ln(S_0/K)+ (r + \sigma^2/2)T}{\sigma \sqrt{T}},  \\
&&\\
d_2 & =& d_1 - \sigma \sqrt{T}, \\
&&\\
N(x) & = \displaystyle & \frac{1}{\sqrt{2\pi}}\int_{-\infty}^{x} \ dz \ e^{-{z^2}/{2}}.\\
&&\\
\end{array}
\end{equation}
The distribution function (\ref{normal_dist}) follows from a stochastic model for stock prices,
\begin{equation}\label{log_norm}
dS = r S dt + \sigma S dz,
\end{equation}
where $dz$ is a Wiener increment~\cite{winer_note}.
It can be shown it is never optimal to exercise an American call 
option on a non-dividend-paying stock early~\cite{Hull},
~\cite{Bouchaud}; 
therefore Eq.~(\ref{BS_solution}) can also be used to estimate the fair value for this kind of options. 

There are some problems with the expressions for $C$ given by 
Eqs.~(\ref{pricing_risk_neutral})-(\ref{BS_solution}). 
Indeed, one can derive \emph{any} option price from
Eq.~(\ref{pricing_risk_neutral}), using different assumptions about 
the distribution function, $P(S_T)$. To derive
from Eq.~(\ref{pricing_risk_neutral}) a result for $C$ which will even approximately coincide with the real market
price, $C^M$, one must specify a distribution function for \emph{future} stock prices, $P(S_T)$. On the
other hand, the expression given by (\ref{BS_solution}) is (a) too specific, and (b) derived using rather strong restrictions. Namely, 
 the stochastic process Eq.~(\ref{log_norm})  does not account for
correlations of returns, $x$ and, moreover,
  the volatility, $\sigma$, and the interest rate, $r$, are
not well-defined parameters (given the actual data). 
As a result, the expression, $C^{BS}$, often
does not coincide (even approximately) with the corresponding market option price, $C^M$.
Useful approaches have been developed which partially solve the problems mentioned
above.

We shall mention here one analysis which is related to that  presented below. This analysis deals with building ``implied trees"~\cite{Jack_rev}. 
There are
many variations of this approach, but the main idea is based on the solution of the
inverse problem: a search for a stock price model that corresponds to the real market
prices of options, $C^M$. A more restricted problem is to search for a stock price model
that effectively deals with the volatility smile.

 In this case, one starts with the BS formula 
(\ref{BS_solution}), (even for American options) but instead of choosing a fixed volatility,
$\sigma = constant$, one uses the dependence, $\sigma = \sigma(K)$ (volatility smile). To some extent,
this dependence corresponds to the ``real behavior" of the volatility, $\sigma$, if one wants to
use Eq.~(\ref{BS_solution}) as the ``zeroth approximation" for option pricing. Details for building trees
(including implied trees) for stock prices can be found in~\cite{Derman_1, Dupire, Derman_2, Rubinstein}. 

There are still some problems with these trees. For example, the corresponding ``implied" stock prices, $S_T$,
can have ``bad" (negative) probabilities which must be eliminated. A solution to this kind of problem, despite the simplicity of the calibration procedure, 
was proposed in Refs~\cite{Jack,Herwing}.

In this paper, we discuss the inverse problem for Eq.~(\ref{pricing_risk_neutral}) using the 
following approach.
First, using Eq.~(\ref{pricing_risk_neutral}), we build the distribution function for future stocks prices and returns from the
empirical data for the market option prices, $C(K)$. Second, we build the returns distribution
functions  using for $C(K)$ a BS expression, $C^{BS}$, with a volatility, $\sigma = \sigma(K)$, in
the form of the volatility smile. 
In particular, we show that the condition of the absence for 
relative minima in the probability distribution function (PDF)
 of returns, (or elimination of ``bad" probabilities) leads to the condition of  ``adiabaticity" for the volatility smile.
This condition can be introduced in the fitting procedure of the volatility smile to get a probability returns distribution more similar to the actual one. In this way one can avoid to generate
 arbitrage opportunities (negative probabilities) in the option pricing methodology (exotic derivatives) and can get a more reliable estimation of the implied volatility. The latter has a key role in the scenario generation and in value at risk 
(VaR)  estimation and has application in the risk management activities.


\section{The Inverse problem for the stock price distribution function}


In this Section, we derive an explicit expression for the distribution function for the future
stock prices, $P(S_T)$ and for returns $P(x)$. In Eq.~(\ref{pricing_risk_neutral})
the distribution function, $P(S_T)$, can be rather arbitrary but it is natural to assume that $P(S_T)$ 
does not depend on the strike price, $K$. According to Eq.~(\ref{pricing_risk_neutral}), the option price, $C$, 
is expressed explicitly through the strike price, $K$.
Differentiating $C$ in Eq.~(\ref{pricing_risk_neutral}) twice with respect to $K$, we have~\cite{Malz},

\begin{equation}\label{price_distr}
P(S_T ) = e^{rT} \frac{\partial^2 C(K)}{\partial K^2} \biggr|_{K=S_T}.
\end{equation}
In Eq.~(\ref{price_distr}), we indicate only the dependence $C(K)$ in the option prices. In particular,
applying Eq.~(\ref{price_distr}) to $C^{BS}$ given in Eq.~(\ref{BS_solution}) 
we derive a distribution function, $P^{BS}(S_T)$, which we present in the form,
\begin{equation}
\begin{array}{lll}
&&P^{BS}(S_T) \equiv \displaystyle e^{rT} \frac{\partial^2 C^{BS}(K)}{\partial K^2} \biggr|_{K=S_T} 
=   \frac{1}{\sqrt{2 \pi \sigma^2 T} S_T} \times \\
&&\\ 
&& \times \displaystyle \exp \left( -\frac{( \ln(S_T/S_0) - (rT - \sigma^2/2 T))^2}{2\sigma^2 T} \right).
\end{array}
\end{equation}
Analogously, the distribution of returns for the Black-Scholes model is Gaussian, as expected:

\begin{equation}\label{distr_BS}
P^{BS}(x) = \frac{1}{\sqrt{2 \pi \sigma^2 T} } \exp \left[ -\frac{( x + x_0)^2}{2\sigma^2 T} \right],
\end{equation}
where $x_0 = \sigma^2/2T$.

We can try to consider the inverse problem substituting the dependence, $\sigma = \sigma(K)$,
in Eq.~(\ref{BS_solution}) and evaluating the distribution of future stocks price and returns, applying
(\ref{price_distr}). After differentiation we get:

\begin{equation}\label{distr_price_pert}
\begin{array}{lll}
&& P(S_T) =\displaystyle
 \frac{F(S_T;S_0,r,T,\sigma)}{\sqrt{2 \pi \sigma^2 T} S_T} \times
\\
&&\\
&& \displaystyle \times \exp\left[ - \frac{( \ln({S_T}/{S_0}) 
- (rT -x_0))^2}{2\sigma^2 T} \right], 
\end{array}
\end{equation}
where we defined:
\begin{equation}
\begin{array}{ccc}
&& F(S_T;S_0, r, T,\sigma)  = \displaystyle \left[1 + S_T 
\frac{\dot{\sigma}}{\sigma} 
\left(r T - \ln({S_T}/{S_0}) \right)\right]^2 \\
&&\\
&& - \displaystyle \frac{(\dot{\sigma}\sigma T S_T)^2}{4} + 
\dot{\sigma}\sigma T S_T + S_T^2 \sigma \ddot{\sigma} T,\\
&&\\
&& \dot{\sigma} =  
\displaystyle \frac{\partial \sigma}{\partial K} \biggr|_{K = S_T},\\
&&\\
&&\ddot{\sigma} =
\displaystyle  \frac{\partial^2 \sigma}{\partial K^2} \biggr|_{K = S_T}.
\end{array}
\end{equation}
Obviously, we can get the expression for the distribution of returns by a simple change of variable 
\begin{equation}
x \equiv \ln\left (\frac{K}{S_0}\right)\biggr|_{K = S_T} - r T,
\end{equation}
so that,
\begin{equation}\label{distr_ret_pert}
P_\sigma (x) = 
\frac{1}{\sqrt{2 \pi \sigma^2 T} } \exp\left[ - 
\frac{( x  + x_0)^2}{2\sigma^2 T} \right] F(x;  T,\sigma),
\end{equation}
where, in a similar way,  we have defined:
\begin{equation}\label{perturb_fact}
\begin{array}{ccc}
&& F(x; T,\sigma) = \displaystyle
(1 - \frac{\sigma^\prime}{\sigma} x  )^2 - \frac{(\sigma^\prime\sigma T)^2}{4} +  
 \sigma \sigma^{\prime \prime} T,\\
&&\\
&& \sigma^\prime = \displaystyle  \frac{\partial \sigma}{\partial x}, \\
&&\\
&& \sigma^{\prime \prime}  =  \displaystyle \frac{\partial^2 \sigma}{\partial x^2}. 
\end{array}
\end{equation}
From (\ref{perturb_fact}) it is clear that if $\sigma$ is constant 
 (\ref{distr_price_pert}) and  (\ref{distr_ret_pert})
are  the distributions for the standard Black-Scholes model; we will call them  ``zeroth-approximation" 
distributions.
If $\sigma \neq const $, the term $F(x;r, T,\sigma)$ could  ``perturb" the relative zero approximation 
(Gaussian)
giving rise to 
distributions that cannot fit real data. As we will  show here,
it is possible to get distributions with 
relative minima (not observed in real returns distributions) and, in the worst case, negative probability.\\
\begin{figure}[t]
\begin{center}
\includegraphics[scale=0.2]{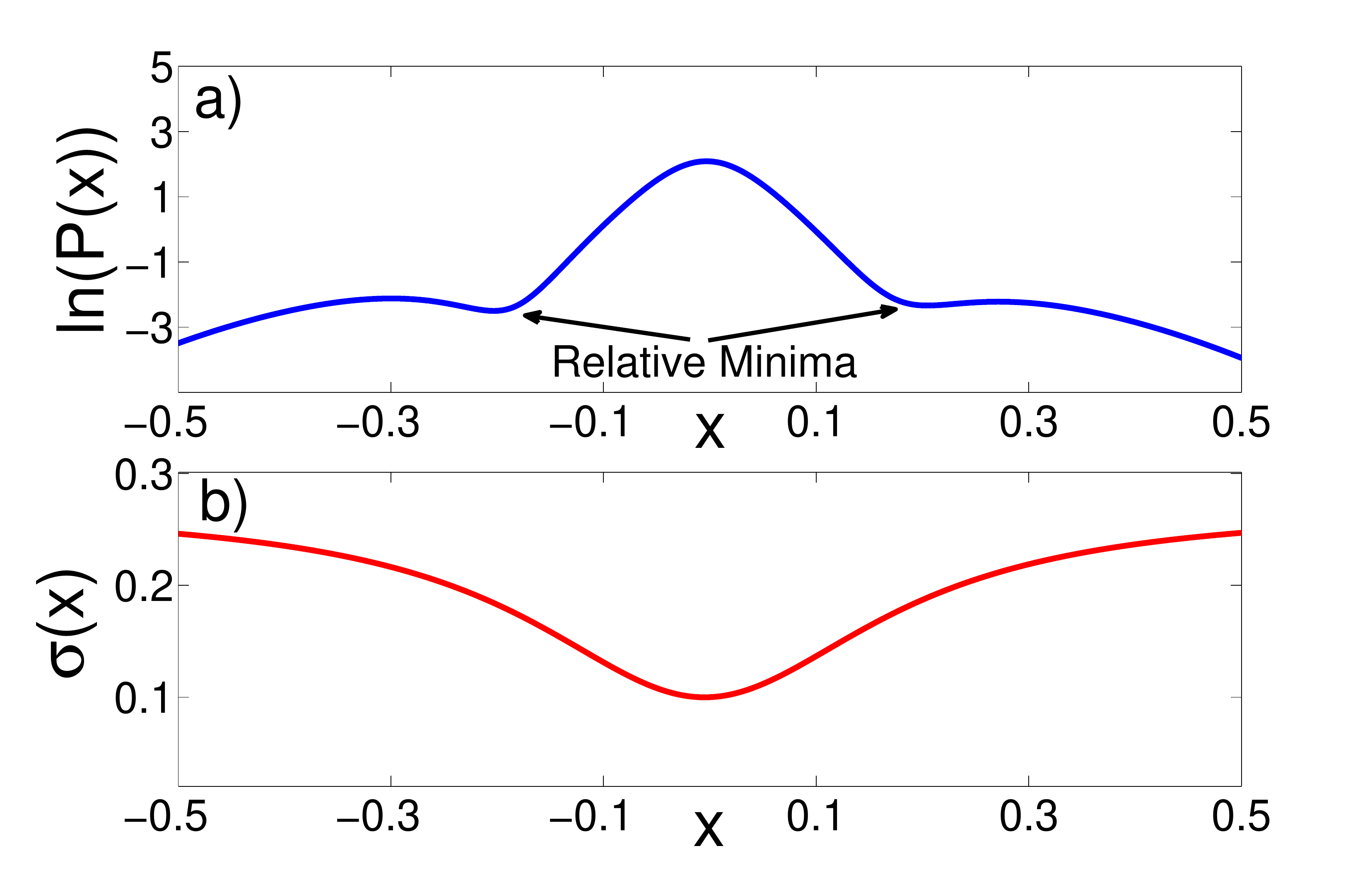}
\end{center}
\caption{a) Volatility smile  given  by Eq.~(\ref{sigma_fit}), with the following parameters: 
$g = 0.1, T=0.5, n = 0.04, \chi = 2.7$. b)
Log-distribution given by Eq.~(\ref{distr_ret_pert}) for  the volatility smile  a).}
\label{pdf_min}
\end{figure}


The rest of the paper is organized as follows. 
In Section \ref{real_data}, we analyze FX market data and we propose a suitable function to fit the volatility smile. We also determine 
the typical range of the parameters we use to perform the fit. We will use these ranges to check our adiabatic condition.
In Section \ref{intuitive_explanation}, we show qualitatively the reason why we have the presence 
of the relative minima in the returns distribution and why an adiabatic approach can be suitable to describe the problem and
used to avoid these ``bad" probabilities.
In Section \ref{results}, we show our numerical and theoretical results about the
 adiabaticity parameter. We also show that there is a critical
value of the adiabatic  parameter which can be used to determine whether
 the returns PDF  will have relative minima. We discuss the relation of this critical value to the other
parameters of the fit. 
Finally, in Section \ref{conclusions} we present our conclusions.


\section{The Volatility Smile: Real Market Data}\label{real_data}


Typically, traders on option markets and practitioners consider the BS model
 as a zeroth order approximation that takes into 
account the main features of options prices. To get a pricing closer to the actual data, they consider the volatility as a
parameter that can be adjusted considering the inverse problem given by Eq.~(\ref{BS_solution}) and the
 real price of
call and put options. In this way  a more reliable value of the 
volatility (\emph{implied volatility}) can be obtained and
it can be used to price more complex options for which analytical solutions are not available. 
The value of the implied
volatility depends on the value of the strike, $K$, in a well-known 
characteristic curve 
 called  the \emph{smile volatility} 
(typically for foreign currency options) whose shape is approximately parabolic and 
symmetric, or \emph{skew volatility}
(typically for equity options) when asymmetric effects dominate~\cite{Risk, Tompkins, Toft, Campa}.

An intuitive explanation of this shape can be found if an actual returns distribution is considered.
 In fact, it is
well known that the tails of the returns PDF are not Gaussian but exhibit a 
power law decay (fat-tails)~\cite{Sorn, Stan}. On the contrary,  BS model
assumes that the PDF
 of returns is Gaussian thus underestimating the actual probability of rare events. To compensate for this model
deficiency, one has to consider the greater implied volatility for strike out of the money then for strike 
at the money. 

In this paper we focus our attention on the volatility smile of foreign currency options and we neglect the skew effect~\cite{Kirch}. 
To perform our analysis we consider the volatility smile as a function of the $\Delta$ of the 
option (defined by Eq.~({\ref{delta})), the time to maturity, $T$, and  the currency 
considered. We consider specific days, for which volatility is not affected by the skew effect, and we use Bloomberg as data provider.
In the BS model, the $\Delta$  of a call options is defined as:
\begin{equation}
\Delta = \frac{\partial C}{\partial S_T} = \displaystyle \frac{ 1}{\sqrt{2 \pi}} \int_{-\infty}^{d_1}\ dz \ 
e^{-{z^2}/{2} }.
\label{delta}
\end{equation}
Inverting this relation is possible to get an expression for $x$:
\begin{equation}\label{Delta_to_x}
x = \sigma^2/2 T - \sigma \sqrt{T} \Phi^{-1} (\Delta),
\end{equation}
where $\Phi^{-1}(x)$ is the inverse of the error function:
\begin{equation}
\Phi(x) = \frac{1}{\sqrt{2\pi}} \int_{-\infty}^{x} e^{-{t^2}/2} \ dt
\end{equation}
\begin{figure}[t]
\begin{center}
\includegraphics[scale=0.2]{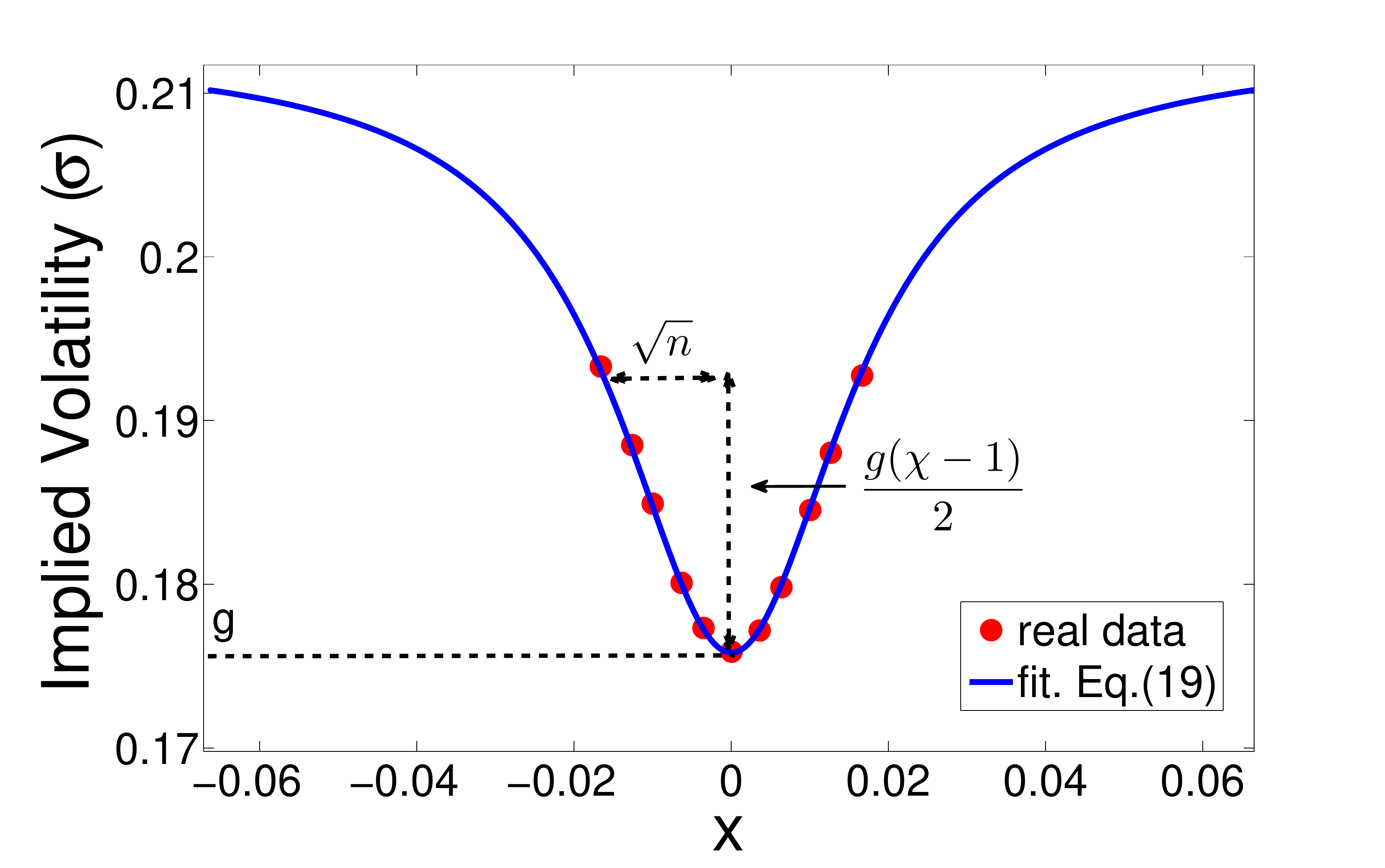}
\end{center}
\caption{Typical volatility smile and the 
relative fit obtained with (\ref{sigma_fit}). 
The parameters of the fit are: $g= 0.1758(5) $, $\chi=1.20(9)
$, $n=0.00030(9) $.
 We get the real data using Bloomberg provider and they refer to the AUDUSD currency with
time to maturity $T=1/365$ years.} 
\label{figfitsmile}
\end{figure}
In Fig.~\ref{figfitsmile}
we show an example of volatility smile in terms of our variables and a 
suitable fit given by the function:
\begin{equation}\label{sigma_fit}
\sigma(x) = g\left[ 1 + (\chi-1)  \frac{  (x + g^2T/2)^2}{(x + g^2T/2)^2 + n}\right]
\end{equation}
where $g, \chi, n$ are fitting parameters. In this case, $g$ represents the minimum of the volatility smile, $\sqrt{n}$ is the half width at the half height, while 
$g(\chi - 1)$ represents
the height of the smile. In particular $\chi$ is the ratio 
between the limiting value of $\sigma$ and $g$ as $x$ approches $\infty$.
In this way the variation of $\sigma$ is bounded between $g$ and $g\chi$.
In the light of the intuitive explanation of the volatility smile proposed above 
and since from~(\ref{distr_BS}) it follows that the average value, 
\begin{equation}
\langle x \rangle  = -\frac{\sigma^2 T}{2},
\end{equation}
one expects that the minimum of the implied volatility occur at $ x = -g^2T/2 $ 
as required by our fitting function. 

Repeating many times the interpolation procedure considering different values for $T$ and currencies, we can
determine typical 
parameters that can fit a wide range of volatility smiles; in the following we will use this 
information to check our results. 

Let us  notice that the following relation between  $n, g$ and $T$ holds:

\begin{equation}\label{n_scaling_rule}
n \propto T g^2,
\end{equation}
as shown in Fig.~\ref{n_scaling}.
\begin{figure}[t]
\begin{center}
\includegraphics[scale=0.2]{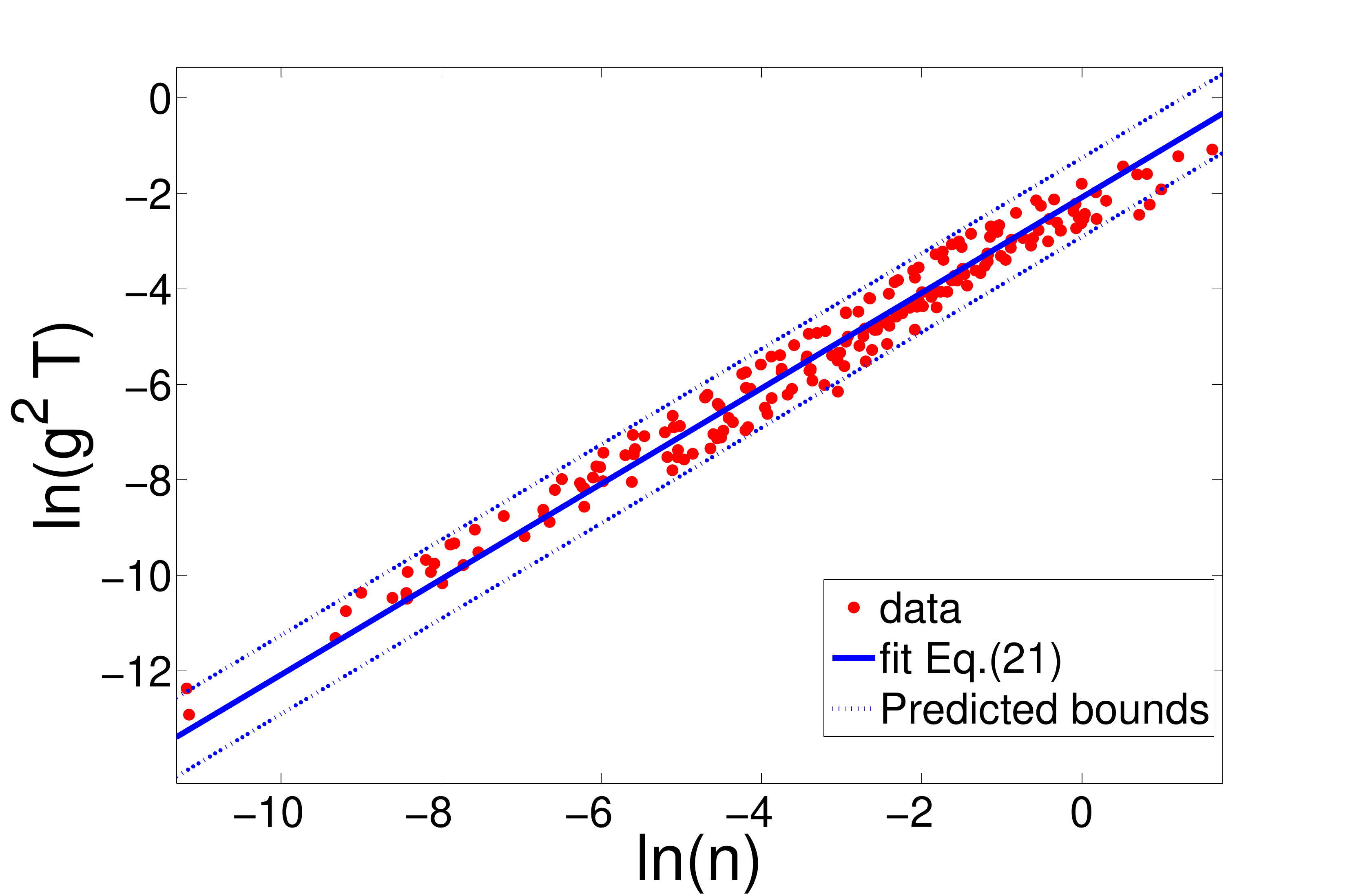}
\end{center}
\caption{Relation between the parameters $n, g, T$. We fit 72 volatility symmetrical smiles (Bloomberg) 
considering different currency (EURUSD, AUDUSD, EURCHF, EURGBP, EURJPY, GBPJPY, GBPUSD, USDCAD, USDCHF, USDJPY)  and time to maturity (1 day, 1-3 weeks, 1, 2, 3, 4, 6, 9 months, 1, 1.5, 2, 3, 4, 5 years)
with the function (\ref{sigma_fit}). We dowload the data on 21/10/2009 and on 01/02/2010. We also show the best linear fit
$\ln (g^2 T) = \ln (n) + c $, where $c =  -1.95(12)$.} 
\label{n_scaling}
\end{figure}
This gives a scaling rule that can be used to determine the range of $n$, fixing $T$ and $g$.


\section{First Approximation of the Volatility Smile: the Squared Well}\label{intuitive_explanation}


In this Section we show qualitatively the reason why there is a relative minimum in the returns distribution 
and why an adiabatic approach can describe the problem of avoiding these ``bad" probabilities.
To keep the problem simple, we consider, as a first step, a volatility smile modeled by a squared well
 defined as follows:
\begin{equation}
\sigma (x) = \left\{
\begin{array}{llr}
\sigma_1 & {\rm  for}  &   |x| < x_1\\
 \sigma_2     & {\rm otherwise} & 
\end{array} ,
\right.
\end{equation}
where $\sigma_2>\sigma_1$ and $x_1$ are positive constants. See Fig.~\ref{sigma_pdf_sq}a,b.
The distribution functions corresponding to two values of
 $\sigma_i$, $i=1,2$
are shown in Fig.~\ref{sigma_pdf_sq}c,d. 
Indicating as $\pm x_1^c$ the abscissa of the intersections
between the two distributions, it is clear that a sufficient condition for avoiding spurious
minima is $x_1 < x_1^c$.
A rough estimation  of $x_1^c$, ignoring the term, $x_0$, usually small, is
\begin{equation}\label{inter_point}
x^c_1 = \sigma_1 \sqrt{T}  \sqrt{\frac{ 2 \chi^2 \ln \chi}{\chi^2 - 1} },
\end{equation}
where $\chi = \sigma_2/\sigma_1$.
Therefore, a sufficient condition to avoid minima in the PDF is to use, as a fitting
function a square well depending on the parameters $x_1, \sigma_1, \sigma_2$, such that $x_1 < x_1^c$.
\begin{figure}[t]
\begin{center}
\includegraphics[scale=0.25]{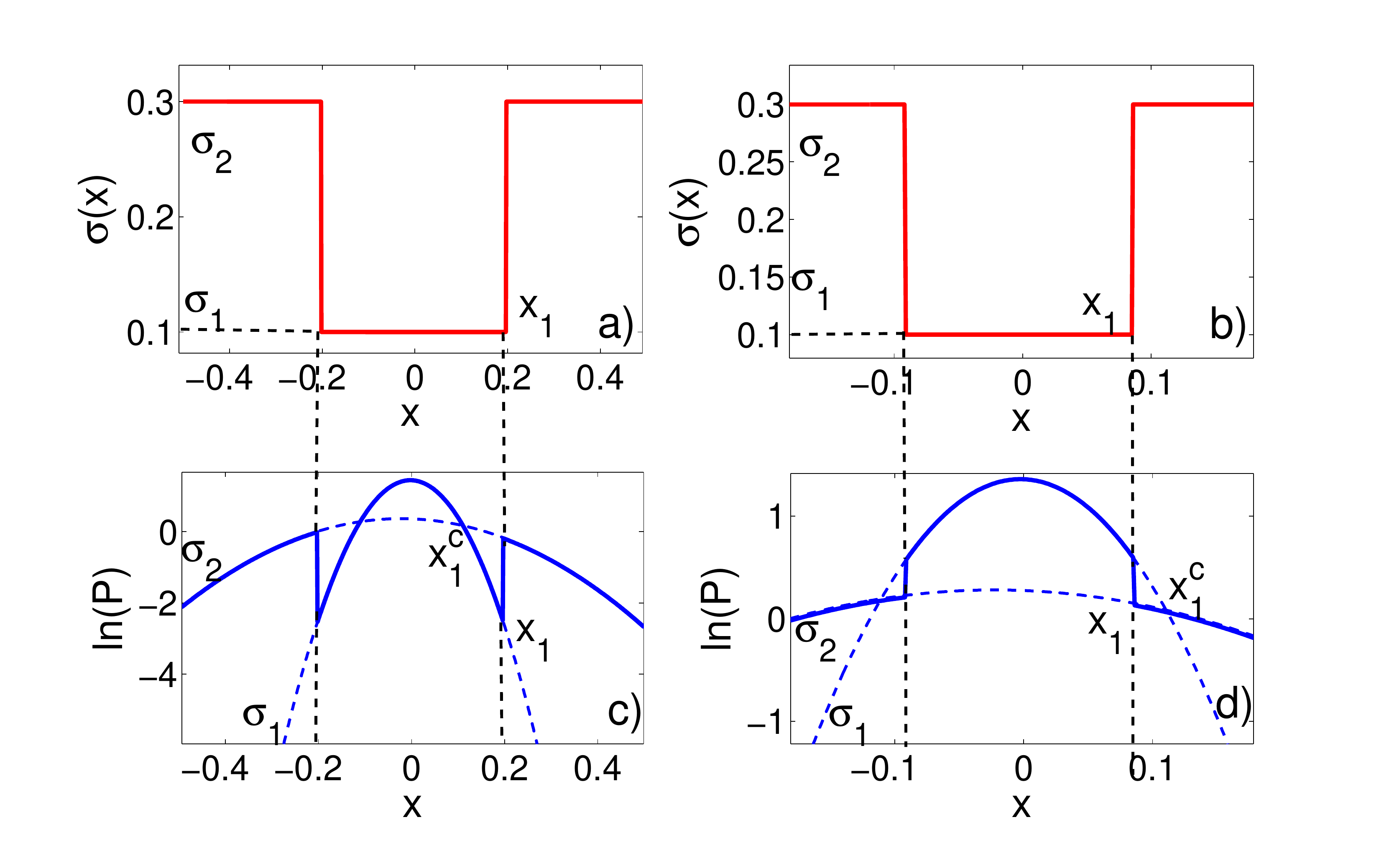}
\end{center}
\caption{a,b) Discontinuous squared well as a fit of 
the volatility smile. c,d) The corresponding PDF
with (c) or without (d) spurious minima.
}
\label{sigma_pdf_sq}
\end{figure}
Therefore, a standard fitting procedure of the volatility smile with a square well, constrained by 
the condition $x_1 < x_1^c$ solves the problem of avoiding spurious minima, even if
is very rough.

If we consider a volatility smile with a continuous variation 
from $\sigma_1$ to $\sigma_2$, we can get, 
instead of a discontinuity, the onset of a relative minimum. The latter can be avoided 
if the variation between $\sigma_1$ and $\sigma_2$ is slow enough so that the connection 
between the two PDF's takes place  keeping constant 
the sign of the first derivative of the distribution during the whole transition. 
So there will be a critical ``speed" of the
transition that will generate zero derivative points which will not correspond 
to the maximum of the distribution. In this case the variable related to the time is 
$x_1$, while $\chi =  \sigma_2/\sigma_1$ can be identified
as a ``distance''. 
To be more precise, one should consider that for  $x < x_1^c $, $P_{\sigma_1} (x) \ge P_{\sigma_2} (x)$,
 so that the effective ``time" should be: $x_1 - x_1^c$.\\ 
We will define an adiabatic transition as one whose passage from $\sigma_1$ to $\sigma_2$
is sufficiently slow, so that  relative minima in the distribution
are not generated. In this same
spirit, we define the critical adiabatic parameter $\chi_c(n,g,T)$ as 
the minimum value of $\chi$ that generates a 
minimum in the distribution, Eq.~(\ref{distr_ret_pert}).
This effect has been shown in  Fig.~(\ref{blu_green_map}),
\begin{figure}[t]
\begin{center}
\includegraphics[scale=0.4]{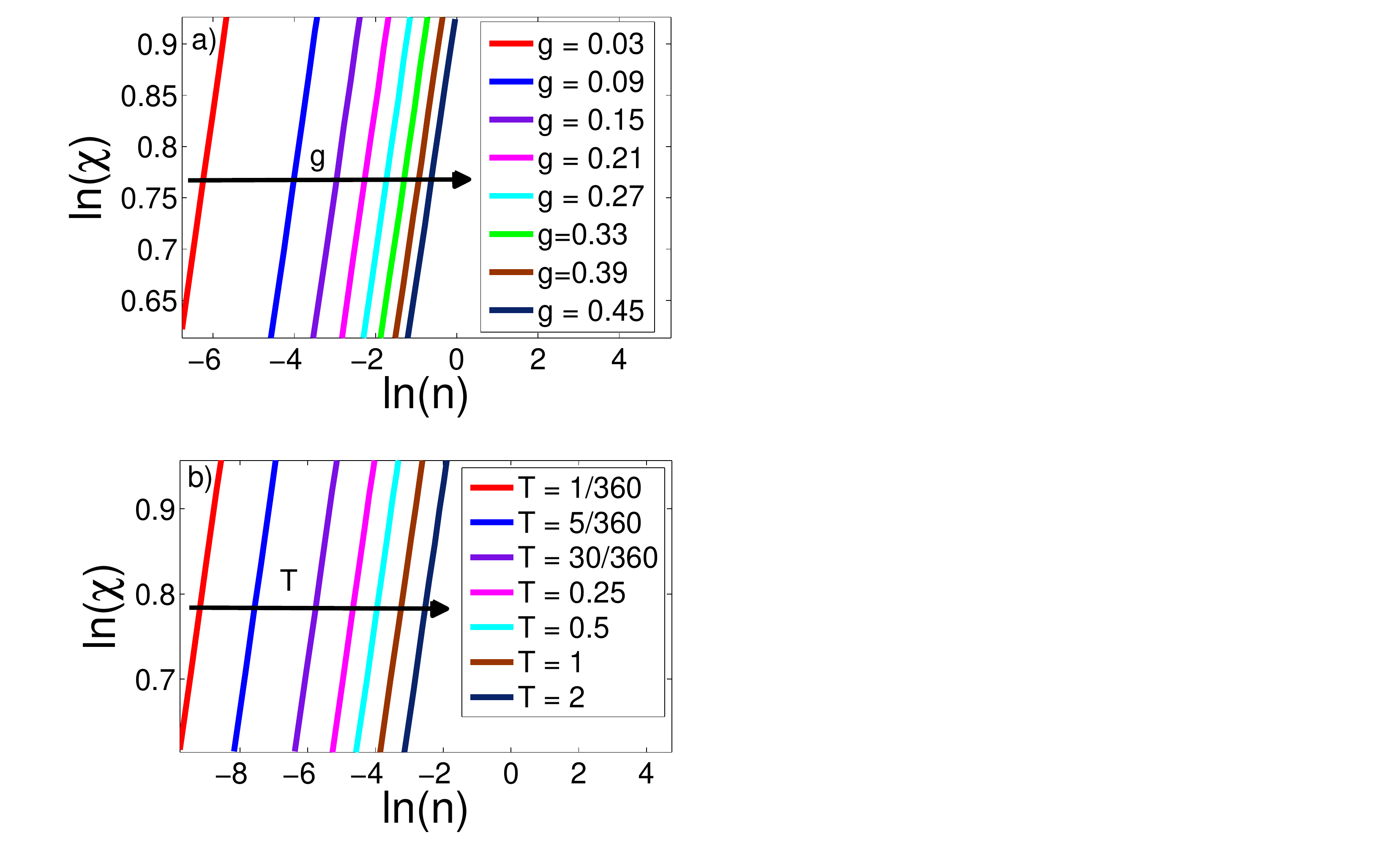}
\end{center}
\caption{a) Critical ratio $\chi_c = \sigma_2/\sigma_1$ as a 
function of $n$ for fixed $T= 0.5$, and different
$g$ as indicated in the legend. 
The arrow indicates the direction of growing $g$. In the region to the right
of the lines
the PDF do not have minima, while in the left hand region  it has.
b) The same as a) but for fixed $g=0.1$ and different $T$ values.
The arrow indicates the direction of growing $T$.}
\label{blu_green_map}
\end{figure}
modeling the volatility smile using (\ref{sigma_fit}), so that $\sigma_1 = g$, 
$\sigma_2 = g\chi$.
We fix the parameters $g, T$  and we vary $n$ and $\chi$, seeking relative minima in the PDF. 
The lines
 in Fig.~\ref{blu_green_map}, obtained respectively at fixed $T$ (a)
and fixed $g$ (b),  divide the plane of parameters into two regions: 
to the left of the lines
the PDF has spurious minima, while this does not happen in the region to the right of it.
It is then clear that for a given set of fixed parameters ($g, T$), there is a relation 
between $\chi$ and $n$ that allow one to obtain a PDF without minima (minima are not observed in real data).
The main goal of this paper is to determine a simple relation that determines whether or not the parameters 
of a volatility smile fit are consistent with real returns distribution and if 
they could give a reliable option pricing.


\section{Numerical and Theoretical Results}\label{results}


In this Section we show our numerical and theoretical results about the 
relation between the set of parameters $n,g,T$
and the critical adiabatic parameter $\chi_c$. 
Using a numerical simulation, we kept fixed $n,g,T$ and we continuously increased
the parameter $\chi$ until we found a zero-derivative point 
for some $x \neq -g^2T/2 $. In this way we could determine 
numerically the critical $\chi_c$. We repeated this approach for a wide range of the parameters values, 
as shown in Table~\ref{tiptab}, 
\begin{table}[htbp]

\begin{center}
\begin{tabular}{|l|c|c|}
\hline
 & \bfseries{min} & \bfseries{max} \\
\hline
\bfseries{g} & $0.03$ & $0.5$ \\
\hline
\bfseries{$\rho$} & $2.5$ & $10$ \\
\hline
\bfseries{T} (years) & $1/365$ & $4$ \\
\hline
\end{tabular}
\end{center}
\caption{Range of the parameters of the numerical simulations}
\label{tiptab}
\end{table}
where we used  the parameter $\rho= n/g^2 T $ instead   of $n$, due to  the scaling relation
(\ref{n_scaling_rule}).
In order to obtain the relation $\chi_c = f_T (n,g)$, we use the following
fit function:
\begin{equation}                    
f_T (n,g) = \alpha \left( \frac{n}{g^2 T} \right)^\beta - 
\gamma \sqrt{T} g \left( \frac{n}{g^2T}\right)^\delta.    
\label{fitf}
\end{equation}  
This has been obtained assuming that the value of the critical parameter $\chi_c$ depends
 on the  rescaled ``time" of the transition  (in our model given by $\rho$). 
We also consider a further
term $\gamma \sqrt{T} g \rho^\delta$ to take
into account  the time correction, $x_c$, as explained in 
Sec.~(\ref{intuitive_explanation}). In this case we make explicit
 the dependence of the time correction on $T$ and $g$ as suggested
 by $\sigma_2$ in Eq.~(\ref{inter_point}).
In Eq.~(\ref{fitf}),  $\alpha, \beta, \gamma, \delta$ are the fitting parameters whose values are
given in  Table~\ref{tab2}.

\begin{table}[htbp]

\begin{center}
\begin{tabular}{|c|c|}
\hline
\bfseries{$\alpha$}  & $1.4373 \pm 0.0002 $  \\
\hline
\bfseries{$\beta$}   & $0.2787  \pm 0.0006 $ \\
\hline
\bfseries{$\gamma$}  & $-0.1738 \pm 0.0002 $ \\
\hline
\bfseries{$\delta$}  & $0.4683 \pm 0.0006  $ \\
\hline
\bfseries{mean squared errors }     & $ 1 \times 10^{-5}  $ \\
\hline
\end{tabular}
\end{center}
\caption{Fitting parameters and relative errors.}
\label{tab2}
\end{table}

In Fig.~\ref{fit_ex} we show the result of our fit for a few selected  values of $g$ and $T$.

\begin{figure}[t]
\begin{center}
\includegraphics[scale=0.2]{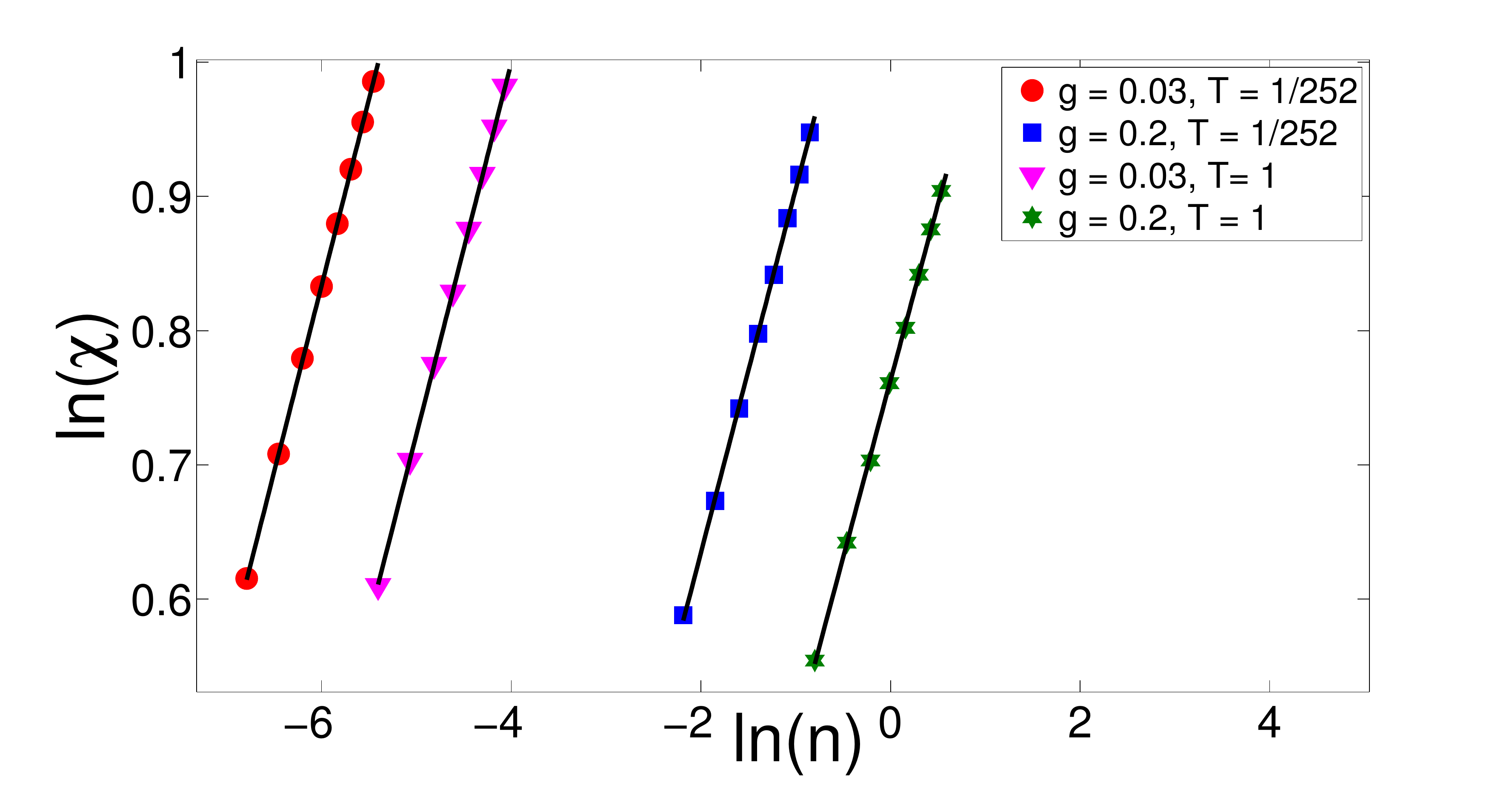}
\end{center}
\caption{Critical adiabatic parameter as a function of $n$  for few selected  pairs
of values  of $(g,T)$ as indicated in the legend.
 The points are numerical data, straight lines are the result of  fitting procedure.}
\label{fit_ex}
\end{figure}

The whole procedure can thus be summarized as follows :
\begin{itemize}
\item The real volatility smile, usually given for a fixed $T$ can be fit by a function dependent
on three parameters $g,n,\chi$, as indicated in Eq.~(\ref{sigma_fit}) and the optimal values
$g^{opt}, n^{opt}, \chi^{opt}$ are returned.
\item The optimal values $g^{opt}, n^{opt}$ are inserted in Eq.~(\ref{fitf}), with 
$\alpha, \beta, \gamma, \delta$ given in Table II and a critical 
$\chi_c = f_T ( n^{opt}, g^{opt}) $ 
obtained.
\item If $\chi^{opt}  < \chi_c$ then we know that  relative 
minima in the PDF do not exist. Otherwise we should perform a fitting procedure to the volatility smile using  Eq.~(\ref{sigma_fit}), constrained by $\chi \leq \chi_c$.   
\end{itemize}

An example of the previous procedure has been shown in Fig.~\ref{ultima}
where the PDF with unwanted minima and the ``corrected '' one is shown
together with the corresponding fitting curve to the volatility smile.
As one can see the price to pay in order to get a smooth PDF is
very small: the two fitting curves for the real volatility smile are
similar, but the PDF has, in the latter case a more realistic behavior.

\begin{figure}[t]
\begin{center}
\includegraphics[scale=0.2]{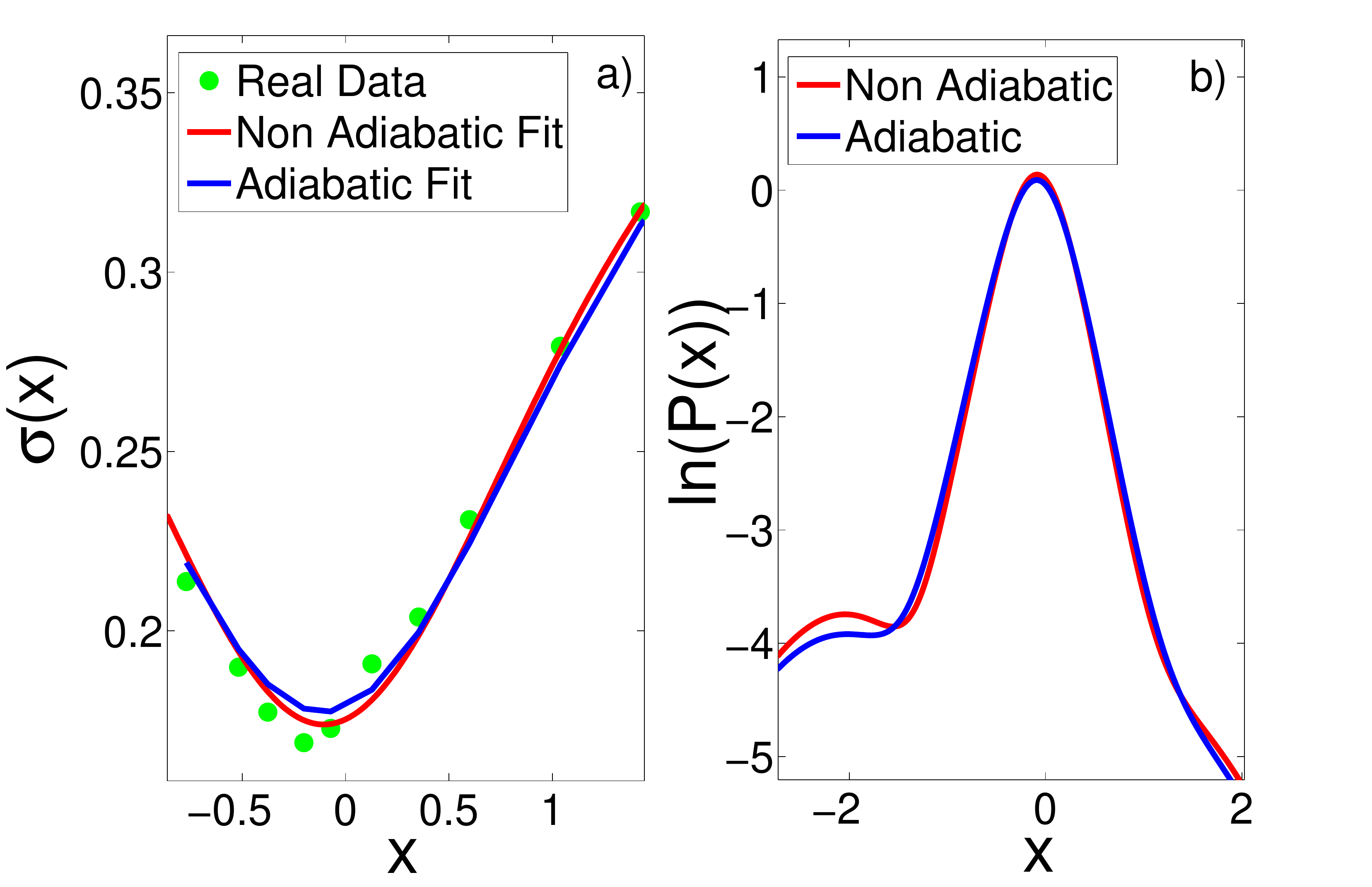}
\end{center}
\caption{a) Volatility smile as a function the returns $x$. Dots indicate real data, red curve 
is the non adiabatic fit, while the blu one represents the adiabatic (constrained) fit. b) PDF of returns for the two curves indicated in a).}
\label{ultima}
\end{figure}

\section{Conclusions}\label{conclusions}

We started from the pricing equation of the Black-Scholes model for an European call
 and we considered a suitable generalization to include the volatility smile effect. 
Then we considered the inverse problem and we analyzed the relative returns distribution, 
Eq.~(\ref{distr_ret_pert}), varying the typical parameters of the volatility smile. We showed
that, for some values of the parameters, it is possible to get relative minima in the returns distribution (bad distribution) that are never observed in real distributions.
 We demonstrated that bad distributions can be eliminated by requiring adiabatic constraints (intuitively 
justified with the example of the squared well) on the volatility smile and we gave a numerical formula to determine the value of the adiabatic critical parameter, $\chi_c$. In this way we provide an easy-to-use tool to determine if a volatility smile fit is consistent with the model hypothesis ($P(x) > 0$) and if it can generate a suitable returns distribution. A reliable estimate of the implied volatility has application in the risk management activities and  in the pricing of exotic derivatives, where, in general, the implied volatility is an input of more complex models.

\section{Acknowledgements}
The work by GPB was carried out under the auspices of the NNSA of the U. S. DOE at LANL under Contract No. DEAC52-06NA2539.


\begin{thebibliography}{}
\bibitem{winer_note} The behavior of the Wiener process can be described on the finite time-interval, $\Delta t$, as $\Delta z  = \gamma \sqrt{\Delta t}$,where the random variable, $\gamma$, has the distribution function: $g(\gamma) = 1/\sqrt{2 \pi} \exp(-\gamma^2/2)$.
\end{thebibliography}
\end{document}